\documentclass[twocolumn,showpacs,aps,prl,superscriptaddress]{revtex4}

\usepackage{graphicx}
\usepackage{dcolumn}
\usepackage{amsmath}
\usepackage{epsfig}

\RequirePackage{xspace}

\usepackage{relsize}
\def\babar{\mbox{\slshape B\kern-0.1em{\smaller A}\kern-0.1em
    B\kern-0.1em{\smaller A\kern-0.2em R}}}

\def\electron   {\ensuremath{e}\xspace}
\def\en         {\ensuremath{e^-}\xspace}   
\def\ep         {\ensuremath{e^+}\xspace}
 
\def\epem       {\ensuremath{e^+e^-}\xspace}

\def\ellm       {\ensuremath{\ell^-}\xspace}
\def\ellp       {\ensuremath{\ell^+}\xspace}

\def\q     {\ensuremath{q}\xspace}

\def\qqbar {\ensuremath{q\overline q}\xspace}
\def\u     {\ensuremath{u}\xspace}

\def\d     {\ensuremath{d}\xspace}

\def\s     {\ensuremath{s}\xspace}

\def\c     {\ensuremath{c}\xspace}

\def\piz   {\ensuremath{\pi^0}\xspace}

\def\Kbar  {\kern 0.2em\overline{\kern -0.2em K}{}\xspace}

\def\Kz    {\ensuremath{K^0}\xspace}
\def\Kzb   {\ensuremath{\Kbar^0}\xspace}
\def\KzKzb {\ensuremath{\Kz \kern -0.16em \Kzb}\xspace}
\def\Kp    {\ensuremath{K^+}\xspace}
\def\Km    {\ensuremath{K^-}\xspace}

\def\KpKm  {\ensuremath{\Kp \kern -0.16em \Km}\xspace}
\def\KS    {\ensuremath{K^0_{\scriptscriptstyle S}}\xspace}

\def\Dbar    {\kern 0.2em\overline{\kern -0.2em D}{}\xspace}

\def\Dz      {\ensuremath{D^0}\xspace}
\def\Dzb     {\ensuremath{\Dbar^0}\xspace}
\def\DzDzb   {\ensuremath{\Dz {\kern -0.16em \Dzb}}\xspace}
\def\Dp      {\ensuremath{D^+}\xspace}
\def\Dm      {\ensuremath{D^-}\xspace}

\def\DpDm    {\ensuremath{\Dp {\kern -0.16em \Dm}}\xspace}

\def\B       {\ensuremath{B}\xspace}
\def\Bbar    {\kern 0.18em\overline{\kern -0.18em B}{}\xspace}

\def\BB      {\ensuremath{B\Bbar}\xspace} 
\def\Bz      {\ensuremath{B^0}\xspace}
\def\Bzb     {\ensuremath{\Bbar^0}\xspace}
\def\BzBzb   {\ensuremath{\Bz {\kern -0.16em \Bzb}}\xspace}
\def\Bu      {\ensuremath{B^+}\xspace}
\def\Bub     {\ensuremath{B^-}\xspace}

\def\BpBm    {\ensuremath{\Bu {\kern -0.16em \Bub}}\xspace}

\def\BorBbar    {\kern 0.18em\optbar{\kern -0.18em B}{}\xspace}
\def\DorDbar    {\kern 0.18em\optbar{\kern -0.18em D}{}\xspace}
\def\KorKbar    {\kern 0.18em\optbar{\kern -0.18em K}{}\xspace}

\def\jpsi     {\ensuremath{{J\mskip -3mu/\mskip -2mu\psi\mskip 2mu}}\xspace}

\mathchardef\Upsilon="7107
\def\Y#1S{\ensuremath{\Upsilon{(#1S)}}\xspace}

\def\FourS {\Y4S}

\mathchardef\Deltares="7101
\mathchardef\Xi="7104
\mathchardef\Lambda="7103
\mathchardef\Sigma="7106
\mathchardef\Omega="710A

\def\Deltabar{\kern 0.25em\overline{\kern -0.25em \Deltares}{}\xspace}
\def\Lbar{\kern 0.2em\overline{\kern -0.2em\Lambda\kern 0.05em}\kern-0.05em{}\xspace}
\def\Sigbar{\kern 0.2em\overline{\kern -0.2em \Sigma}{}\xspace}
\def\Xibar{\kern 0.2em\overline{\kern -0.2em \Xi}{}\xspace}
\def\Obar{\kern 0.2em\overline{\kern -0.2em \Omega}{}\xspace}
\def\Nbar{\kern 0.2em\overline{\kern -0.2em N}{}\xspace}
\def\Xb{\kern 0.2em\overline{\kern -0.2em X}{}\xspace}

\def\mes        {\mbox{$m_{\rm ES}$}\xspace}

\def\DeltaE     {\mbox{$\Delta E$}\xspace}

\newcommand{\tev}{\ensuremath{\mathrm{\,Te\kern -0.1em V}}\xspace}
\newcommand{\gev}{\ensuremath{\mathrm{\,Ge\kern -0.1em V}}\xspace}
\newcommand{\mev}{\ensuremath{\mathrm{\,Me\kern -0.1em V}}\xspace}
\newcommand{\kev}{\ensuremath{\mathrm{\,ke\kern -0.1em V}}\xspace}
\newcommand{\ev}{\ensuremath{\mathrm{\,e\kern -0.1em V}}\xspace}
\newcommand{\gevc}{\ensuremath{{\mathrm{\,Ge\kern -0.1em V\!/}c}}\xspace}
\newcommand{\mevc}{\ensuremath{{\mathrm{\,Me\kern -0.1em V\!/}c}}\xspace}
\newcommand{\gevcc}{\ensuremath{{\mathrm{\,Ge\kern -0.1em V\!/}c^2}}\xspace}
\newcommand{\mevcc}{\ensuremath{{\mathrm{\,Me\kern -0.1em V\!/}c^2}}\xspace}

\def\mus  {\ensuremath{\rm \,\mus}\xspace}

\def\ps   {\ensuremath{\rm \,ps}\xspace}

\def\mus        {\ensuremath{\,\mu{\rm s}}\xspace}    
\def\ps         {\ensuremath{{\rm \,ps}}\xspace}  

\def\to                 {\ensuremath{\rightarrow}\xspace}

\newcommand{\stat}{\ensuremath{\mathrm{(stat)}}\xspace}
\newcommand{\syst}{\ensuremath{\mathrm{(syst)}}\xspace}

\def\pep2{PEP-II}

\def\gsim{{~\raise.15em\hbox{$>$}\kern-.85em
          \lower.35em\hbox{$\sim$}~}\xspace}
\def\lsim{{~\raise.15em\hbox{$<$}\kern-.85em
          \lower.35em\hbox{$\sim$}~}\xspace}

\def\CP                {\ensuremath{C\!P}\xspace}
\def\C       {\ensuremath{C}\xspace}

\def\deltat{\ensuremath{{\rm \Delta}t}\xspace}
\def\deltamd{\ensuremath{{\rm \Delta}m_d}\xspace}

\xspace

\newcommand{\jprlBase}       {Phys.\ Rev.\ Lett.\xspace}
\newcommand{\jprBase}        {Phys.\ Rev.\xspace}
\newcommand{\jplBase}        {Phys.\ Lett.\xspace}
\newcommand{\nimBaseA}       {Nucl.\ Instrum.\ Methods Phys.\ Res., Sect.\ A\xspace}

\newcommand{\npBase}         {Nucl.\ Phys.\xspace}

\newcommand{\jpg}       [1]  {{J.\ Phys.\ {\bf G{\bf #1}}}}

\newcommand{\nima}      [1]  {\nimBaseA~{\bf #1}}

\newcommand{\np}        [1]  {\npBase\ {\bf #1}}
\newcommand{\plb}       [1]  {\jplBase\ B~{\bf #1}}

\newcommand{\jprl}      [1]  {\jprlBase\ {\bf #1}}
\newcommand{\pr}        [1]  {\jprBase\ {\bf #1}}
\newcommand{\jprd}      [1]  {\jprBase\ D~{\bf #1}}

\def\jetset74   {\mbox{\tt Jetset \hspace{-0.5em}7.\hspace{-0.2em}4}\xspace}

\def\sintwob    {\ensuremath{\sin 2 \beta}}

\def\bpmtojpsirho {\ensuremath{\B^\pm \to \jpsi\rho^\pm}}
\def\fish    {\ensuremath{\cal F}}

\def\btojpsirho {\ensuremath{\B^0 \to \jpsi\rhoz}}

\def\btojpsipi {\ensuremath{\B^0 \to \jpsi\piz}}
\def\btojpsiks {\ensuremath{\B^0 \to \jpsi\KS}}

\def\bpmtojpsikstar {\ensuremath{\B^\pm \to \jpsi\:{K}^{*\pm}}}
\def\btojpsikstar{\ensuremath{\B^0 \to {J}/\psi\:{K}^{*0}}}

\def\jpsitoll {\ensuremath{\jpsi\rightarrow\ellp\ellm}}
\def\jpsitoee {\ensuremath{\jpsi\rightarrow\ep\en}}

\def\C {\ensuremath{C}}
\def\S {\ensuremath{S}}

\def\dt   {\ensuremath{\Delta t}}

\def\dz   {\ensuremath{\Delta z}}

\def\d    {\ensuremath{d}}
\def\u    {\ensuremath{u}}
\def\s    {\ensuremath{s}}

\def\bb   {\ensuremath{B\Bbar}}

\def\rhoz {\ensuremath{\rho^0}}

\def\signalyield{\ensuremath{184 \pm 15 \stat}}
\def\fittedS{\ensuremath{-1.23 \pm 0.21 \stat}}
\def\fittedC{\ensuremath{-0.20 \pm 0.19 \stat}}
\def\measuredbf{\ensuremath{(1.69 \pm 0.14 \stat \pm 0.07 \syst)\times 10^{-5}}}
\def\measuredS{\ensuremath{\fittedS \pm 0.04 \syst}}
\def\measuredC{\ensuremath{\fittedC \pm 0.03 \syst}}
\def\rhoSC{\ensuremath{19.7\%}}

\newcommand{\BABARPubYear}    {08}
\newcommand{\BABARPubNumber}  {10}

\def\figurebox#1#2#3{%
    \def\arg{#3}%
    \ifx\arg\empty
    {\hfill\vbox{\hsize#2\hrule\hbox to #2{\vrule\hfill\vbox to #1{\hsize#2\vfill}\vrule}\hrule}\hfill}%
    \else
    {\hfill\epsfbox{#3}\hfill}%
    \fi}

\begin{document}

\preprint{\babar-PUB-\BABARPubYear/\BABARPubNumber} 

\begin{flushleft}
\babar-PUB-\BABARPubYear/\BABARPubNumber\\
\end{flushleft}

\title{
{\large \bf
\boldmath Evidence for \CP\ violation in {\boldmath $\btojpsipi$} decays} 
}

%
\author{B.~Aubert}
\author{M.~Bona}
\author{Y.~Karyotakis}
\author{J.~P.~Lees}
\author{V.~Poireau}
\author{E.~Prencipe}
\author{X.~Prudent}
\author{V.~Tisserand}
\affiliation{Laboratoire de Physique des Particules, IN2P3/CNRS et Universit\'e de Savoie, F-74941 Annecy-Le-Vieux, France }
\author{J.~Garra~Tico}
\author{E.~Grauges}
\affiliation{Universitat de Barcelona, Facultat de Fisica, Departament ECM, E-08028 Barcelona, Spain }
% ...
\collaboration{The \babar\ Collaboration}
\noaffiliation

\date{\today}

%
%
\begin{abstract}
We present measurements of the branching fraction and time-dependent \CP\ asymmetries
in $\btojpsipi$ decays based on 466 million
$\FourS\to\bb$ events collected with the \babar\ detector at the SLAC PEP-II
asymmetric-energy \B\ factory.  We measure the \CP\ asymmetry parameters $\S=\measuredS$ 
and $\C=\measuredC$, where the measured value of $\S$ is $4.0$ standard deviations from zero
including systematic uncertainties.
The branching fraction is determined to be ${\cal{B}}(\btojpsipi) = \measuredbf$. 
\end{abstract}


\maketitle


Charge conjugation-parity (\CP) violation in the $B$ meson system has been
established by the \babar~\cite{babar-stwob-prl}
and Belle~\cite{belle-stwob-prl} collaborations.
The Standard Model (SM) of electroweak interactions describes \CP\ violation
as a consequence of a complex phase in the
three-generation Cabibbo-Kobayashi-Maskawa (CKM) quark-mixing
matrix~\cite{ref:CKM}. Measurements of \CP\ asymmetries in
the proper-time distribution of neutral $B$ decays to
\CP\ eigenstates containing a $\jpsi$ and $K^{0}$ meson provide
a precise measurement of $\sintwob$~\cite{BCP}, where
$\beta$ is $\arg \left[\, -V_{\rm cd}^{}V_{\rm cb}^* / V_{\rm td}^{}V_{\rm tb}^*\, \right]$ and
the $V_{\rm ij}$ are CKM matrix elements with $i$, $j$ quark indices. 

The decay $\btojpsipi$ is a Cabibbo-suppressed 
${b \rightarrow c\mskip 2mu \overline c \mskip 2mu d}$ transition
to a \CP-even final state
whose tree amplitude has the same weak phase as the
${b \rightarrow c\mskip 2mu \overline c \mskip 2mu s}$ modes,
{\emph{e.g.}}, the decay $\btojpsiks$. 
The ${b \rightarrow c\mskip 2mu \overline c \mskip 2mu d}$
loop (penguin) amplitudes have different weak phases than the tree amplitude.
If there is a significant penguin amplitude in $\btojpsipi$, then 
the measured values of the \CP asymmetry coefficients \S\ and \C\ 
will differ from the tree level expectations of $-\sintwob$ and 0, 
respectively, and this mode could be sensitive to physics beyond 
the SM~\cite{grossman}.
The coefficient \S\ is related to \CP\ violation in interference between 
amplitudes of direct decay, and decay after mixing, and \C\ is related to direct \CP\ violation.
An additional motivation for measuring \S\ and \C\ from $\btojpsipi$ is that 
they can provide a model-independent constraint on the penguin contamination 
within $\btojpsiks$~\cite{ciuchini}.

The data used in this analysis were collected with the \babar\ detector~\cite{ref:babar} 
at the \pep2\ asymmetric $\epem$ 
storage ring~\cite{ref:pepii}. This represents an integrated luminosity 
of 425 fb$^{-1}$ collected on the $\FourS$ resonance (on-peak),
 which corresponds to (466 $\pm$ 5) million $\BB$ pairs. 
In this letter, we present an update of our previous  
measurements of the branching fraction ${\cal {B}}$ and \CP\ asymmetries 
of $\btojpsipi$~\cite{ref:babarjpsipiz}, which had 
been performed using an integrated luminosity of 232 fb$^{-1}$. 
Belle has also studied this mode and has published a branching
fraction and a time-dependent \CP\-violating asymmetry result using
29.4 fb$^{-1}$ and 484.3 fb$^{-1}$ of integrated luminosity, respectively~\cite{ref:bellebf,ref:bellecp}. 

We reconstruct $\btojpsipi$ decays from combinations
 of $\jpsitoll$ ($\ell$ = $\electron$, $\mu$) and $\piz\rightarrow\gamma\gamma$ candidates.
A detailed description of the charged particle reconstruction and identification
can be found elsewhere~\cite{ref:bigprd}. For the $\jpsitoee$ ($\mu^+\mu^-$) channel, 
the invariant mass of the lepton pair is required to lie between $3.06$ and $3.12 \gevcc$ 
($3.07$ and $3.13 \gevcc$). Each lepton candidate must be consistent with the 
electron (muon) signature in the detector. We form $\piz\to\gamma\gamma$ candidates from clusters in 
the electromagnetic calorimeter with an invariant mass, $m_{\gamma\gamma}$, 
satisfying $100 < m_{\gamma\gamma} < 160$ {\mevcc}.
These clusters are required to be isolated from any charged tracks, carry
a minimum energy of 30{\mev}, and have a lateral energy distribution 
 consistent with that of a photon. Each $\piz$ candidate is required to
have a minimum energy of 200{\mev} and is constrained
to the nominal mass~\cite{ref:pdg2006}. 

We use two kinematic variables, $\mes$ and \DeltaE, in order to isolate the signal:
$\mes=\sqrt{(s/2 + {\mathbf {p}}_i\cdot {\mathbf {p}}_B)^2/E_i^2- {\mathbf {p}}_B^2}$ is the beam-energy substituted mass and
$\DeltaE = E_B^* - \sqrt{s}/2$ is the difference between the \B-candidate energy 
and the beam energy. Here the $\btojpsipi$ candidate ($B_{\mathrm{rec}}$) momentum
${\mathbf {p}_B}$ and four-momentum of the initial state $(E_i, {\mathbf
{p}_i})$ are defined in the laboratory frame, $E_B^*$ is the $B_{\mathrm{rec}}$ energy in 
the center-of-mass (CM) frame, and $\sqrt{s}/2$ is the beam energy in the CM frame. We require 
$\mes > 5.2 \gevcc$ and $-0.1 < \DeltaE < 0.3 \gev$.  The asymmetric
$\DeltaE$ cut is used in order to reduce background from $B$ meson decays to 
final states including a \jpsi\ meson, where one or more of the particles 
in the final state is not reconstructed as part of $B_{\mathrm{rec}}$.

A significant source of background is from \epem\to\qqbar ($\q = \u,\d,\s,\c$) continuum 
events.  We combine several kinematic and topological variables into a Fisher discriminant (\fish)
to provide additional separation between signal and continuum. 
The three variables $\cos$($\theta_{H}$), $L_0$, and $L_2$ are inputs to \fish,
where $\theta_{H}$ is the angle between the positively charged lepton
and the $\B$ candidate momenta in the $\jpsi$ rest frame.
The variables $L_0$ and $L_2$ are the zeroth- and second-order moments;
$L_0 = \sum_i |{\bf p}^{\rm *}_i|$ and 
$L_2 = \sum_i |{\bf p}^{\rm *}_i| \hspace{0.5mm} {(3 \cos^2\theta_i - 1)}/2$,
where ${\bf p}^{\rm *}_i$ are the CM momenta of the tracks and neutral
calorimeter clusters that are not associated with the signal candidate. The
$\theta_i$ are the angles between ${\bf p}^{\rm *}_i$ and the thrust axis of
the signal candidate.  We use data collected 40 \mev\ below
the $\FourS$ resonance to model background from continuum events, and 
signal Monte Carlo (MC) simulated data 
to calculate the coefficients used in \fish.


We use multivariate algorithms to identify signatures that determine (tag)
the flavor of the decay of the other \B in the event ($\B_{\mathrm{tag}}$) to be either 
a \Bz or \Bzb. The flavor tagging algorithm has seven mutually exclusive categories 
of events and is described in detail elsewhere~\cite{ref:babar2007}.
The total effective tagging efficiency of this algorithm is 
given by $\sum_{i}\epsilon_i(1-2\omega_i)^2=(30.5 \pm 0.4)\%$, 
where $\epsilon_i$ is the efficiency of a tag, $\omega_i$ is 
the probability of mis-identifying a tag,
and  $i$ runs over the seven tag categories.

The decay rate $f_+$ ($f_-$) of neutral decays to a \CP eigenstate, 
when $B_{\mathrm{tag}}$ is a  \Bz (\Bzb), is:
\begin{equation}
f_{\pm}(\dt) = \frac{e^{-\left|\dt\right|/\tau_{\Bz}}}{4\tau_{\Bz}} [1
\pm S\sin(\deltamd\deltat) \mp \C\cos(\deltamd\dt)],  
\label{equation-ff}
\end{equation}
where \dt\ is the difference between the proper decay times of 
the $B_{\mathrm{rec}}$ and $B_{\mathrm{tag}}$ mesons, $\tau_{\Bz}$ = 1.530 $\pm$ 0.009 ps is the \Bz\ lifetime
and \deltamd\ = 0.507 $\pm$ 0.005 ps$^{-1}$ is the \Bz-\Bzb\ oscillation 
angular frequency~\cite{ref:pdg2006}. 
The decay width difference between the \Bz\ mass eigenstates is assumed to be zero. 

The time interval \dt\ is calculated from the measured separation \dz\ between
the decay vertices of $B_{\mathrm{rec}}$ and $B_{\mathrm{tag}}$ along the collision axis ($z$).
The vertex of $B_{\mathrm{rec}}$ is reconstructed from the lepton tracks that come from the $J/\psi$;
the vertex of $B_{\mathrm{tag}}$ is constructed from tracks in the event that do
not belong to $B_{\mathrm{rec}}$, with constraints from the beam spot location
and the $B_{\mathrm{rec}}$ momentum.  We accept events with $|\dt|<20 \ps$ whose
uncertainty $\sigma(\dt)$ is less than $2.5 \ps$.  

After the selection criteria mentioned above are applied, the average
number of candidates per event is approximately 1.1 in data.
The multiple candidates per event result from having more than one choice 
of $\piz$ per event, so we choose the one whose value of $m_{\gamma\gamma}$ is closest to 
the $\piz$ mass reported by the PDG~\cite{ref:pdg2006}.
Overall, the true signal candidate is correctly identified $99.6 \%$ of the time for signal MC simulated data.
After this step, the signal efficiency is $19.3 \%$ and a total of 1120 
events are selected in on-peak data.

In addition to signal and continuum background events, there are also \bb-associated backgrounds present 
in the data.
We consider \B\ backgrounds from the following types of event: (i) \btojpsiks, (ii) $\btojpsikstar$, 
(iii) $\bpmtojpsikstar$, (iv) $\bpmtojpsirho$, (v) $\btojpsirho$, (vi) other \B\ decays to final 
states including a $\jpsi$, and (vii) \B\ meson decays to final states including charm mesons.
The yields of these backgrounds
are fixed to expectations (16.2, 9.4, 8.8, 2.3, 0.3, 79.4, and 60.4 events, respectively), 
using branching ratios from world averages~\cite{ref:hfag}.  
We allow these to vary in turn when evaluating systematic uncertainties. 
Backgrounds from other \B\ decays are small, and have been neglected.

The signal yield, \S, and \C\ are simultaneously extracted from an unbinned extended maximum-likelihood (ML)
fit to the on-peak data sample, where the discriminating variables used in the fit are 
\mes, \DeltaE, \fish\ and \dt. For each candidate-type (signal, 
continuum, and the aforementioned \B\ backgrounds) 
we construct a probability density function (PDF) that is the product of PDFs
in each of these variables, assuming that they are uncorrelated.  These
combined PDFs are used in the fit to the data sample.
The continuum-background \mes, \DeltaE, \fish, and \dt\ PDF parameters are 
floated in the final fit to the data. For all other types the
PDF parameters are extracted from high-statistics MC samples. 
The \mes\ distributions for signal and $\btojpsiks$ events peak at the \B\ mass, and
are described by a Gaussian with a low side exponential tail (GE).  The \mes\ PDFs for all other backgrounds are described by ARGUS functions~\cite{ref:argus}.
The signal \DeltaE\ distribution is described by a sum of a GE distribution 
and a second order polynomial. We use a smoothed histogram of MC simulated
data to describe the \DeltaE\ PDFs for \btojpsiks, \bpmtojpsirho, and  \B\ meson 
decays to final states including charm mesons, and second order polynomials 
for the \DeltaE\ PDFs of all other backgrounds. 
We parameterize the \fish\ distribution for signal and continuum
events using the sum of a Gaussian and a Gaussian with different widths above, and below the mean.
The \fish\ distributions for all other background PDFs are Gaussians.
The signal \deltat\ distribution is described by Eq. (\ref{equation-ff}) convolved with 
three Gaussians (core, tail, outliers) which takes into account $\sigma(\deltat)$ from 
the vertex fit, and tagging dilution. The resolution is parameterized using a
large sample of fully reconstructed hadronic \B\ decays~\cite{ref:babar2007}.
The nominal \deltat\ distribution for the \B\ backgrounds is the same as for signal, 
except for inclusive \B\ and $\jpsi K^{*0}$ backgrounds, which use an effective lifetime 
determined from MC samples of $1.1\ps$.
The continuum background \deltat\ distribution is described by the sum of three
Gaussian distributions.   The \dt\ PDF parameters depend on the flavor tag
category.   The signal yield is fitted using known tag efficiencies
listed in Ref.~\cite{ref:babar2007} for each tag category.   The continuum
yields for the seven tagging categories are allowed to vary in the ML fit, and
the fractions of \B\ background events in each category are
determined from MC samples.

\begin{figure}[ht]
\begin{center}
\resizebox{8.7cm}{!}{
\includegraphics{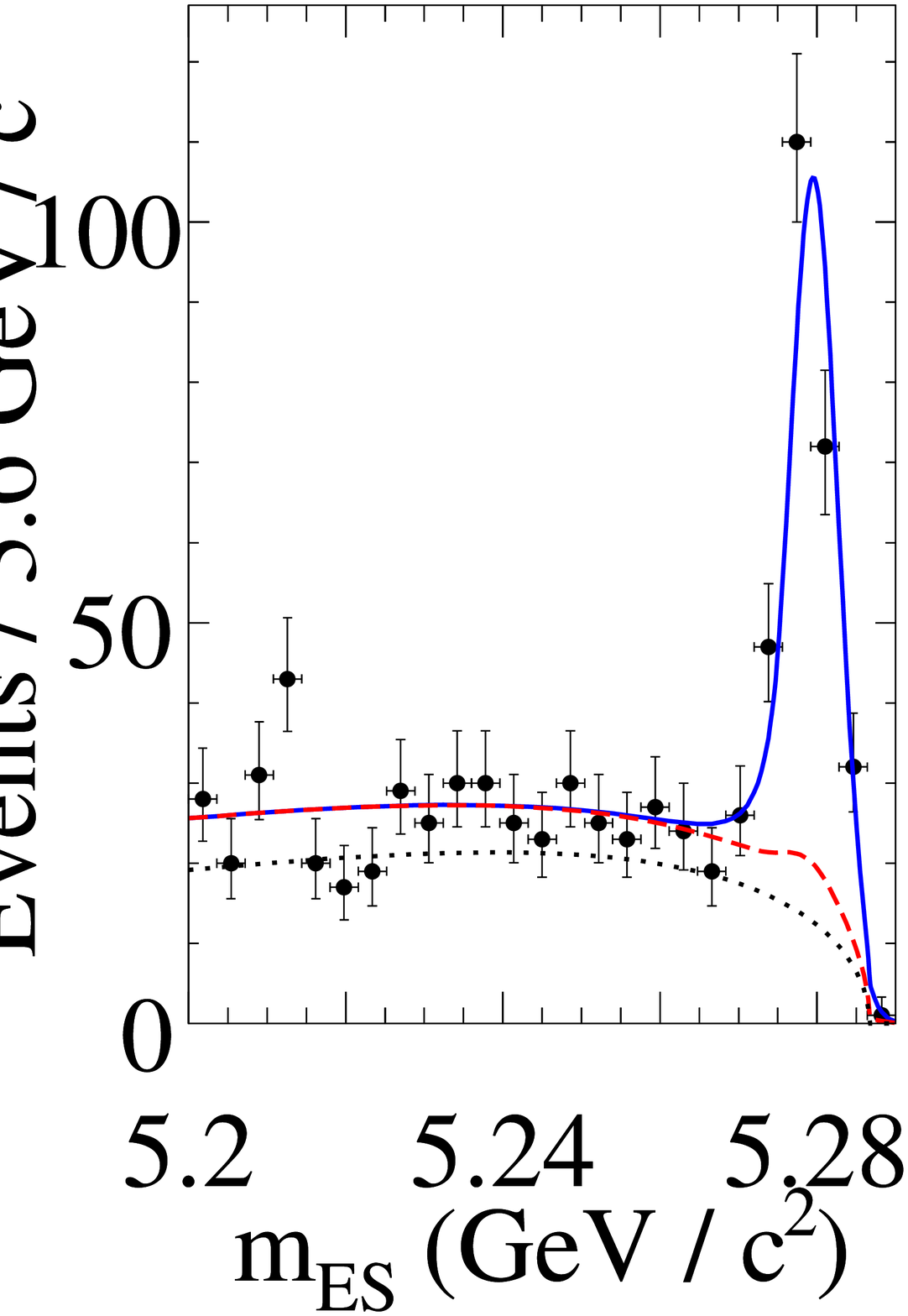}
\includegraphics{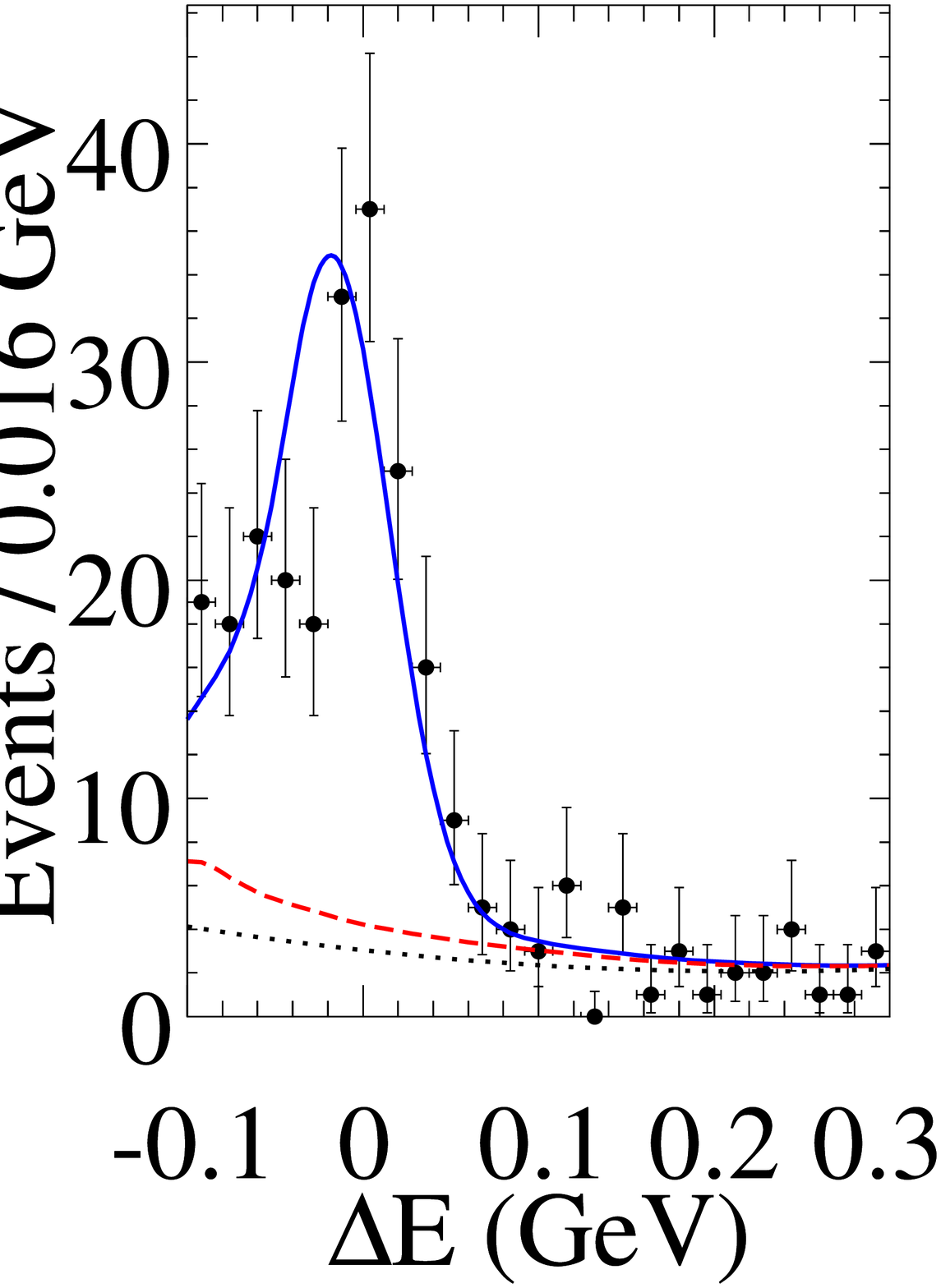}
\includegraphics{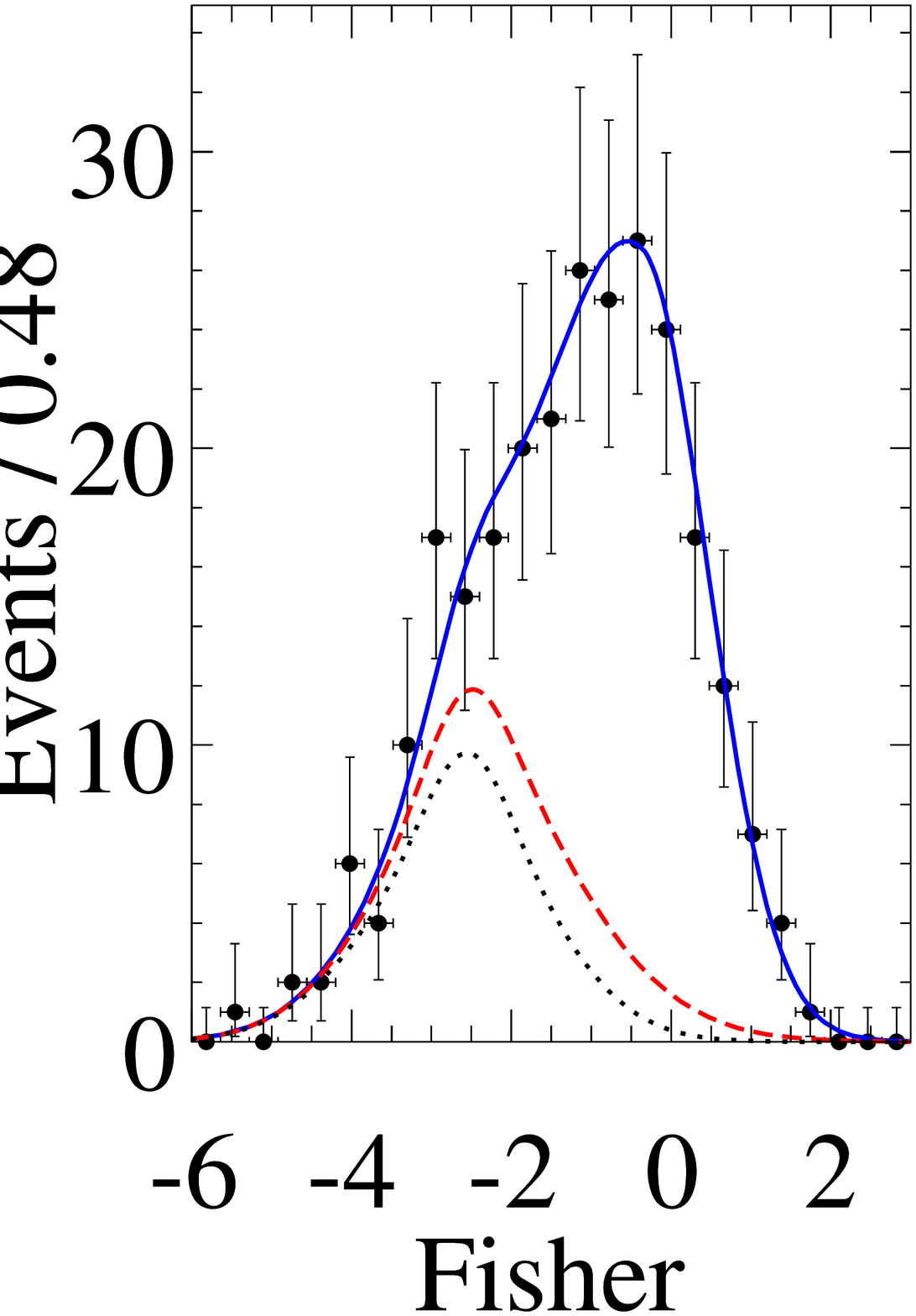}
}
\caption{Signal enhanced distribution of (left) \mes, 
(middle) \DeltaE, and (right) \fish\ for the 
data (points), sum of signal and backgrounds (solid line), sum of backgrounds
(dashed line), and the continuum background (dotted line).}
\label{fig:projection}
\end{center}
\end{figure}

\begin{figure}[ht]
\begin{center}
\includegraphics[height=6.0cm]{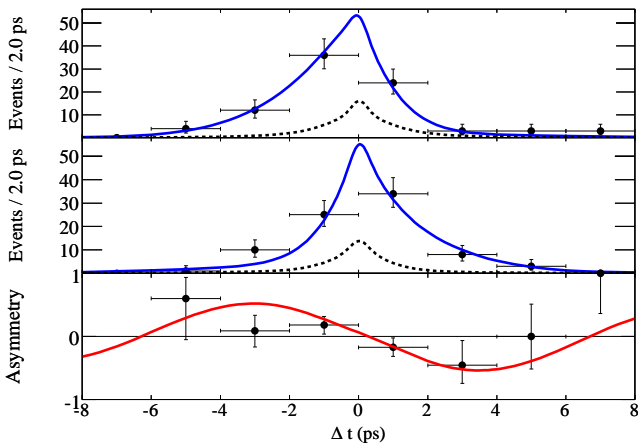}
\caption{The \deltat\ distribution for a sample of signal enhanced events 
tagged as $\Bz$ (top) and $\Bzb$ (middle), where dotted lines are the sum 
of backgrounds and solid lines are the sum of signal and backgrounds.  
The time-dependent \CP asymmetry (see text) is also shown (bottom), where the curve represents 
the measured asymmetry.}
\label{fig:asym}
\end{center}
\end{figure}

After performing tests on the fitting procedure as described in 
Ref.~\cite{babar:rhorhoprd}, we fit the data.
The results, corrected for fit bias, are 
$\signalyield$ signal events, \S\ = \fittedS\ and 
\C = \fittedC.  Figure~\ref{fig:projection} shows distributions of \mes, \DeltaE, and \fish\ for the 
data, where the signal is enhanced by 
selecting $\DeltaE < 0.1 \gev$ for the \mes\ distribution, and $\mes>5.275\gevcc$ for the other distributions. 
These requirements have a relative signal efficiency of 98.8\% (92.3\%) and background efficiency 
of 64\% (10.4\%) for \mes (\DeltaE\ and \fish).
Figure~\ref{fig:asym} shows the \deltat\ distributions for signal \Bz and \Bzb
tagged events. The signal is enhanced by excluding events from the tagging category 
with the largest value of $\omega$, 
and by requiring  $\mes>5.275\gevcc$ and $\DeltaE < 0.1 \gev$.
These requirements have a relative efficiency of 70.0\% (4.4\%) for 
signal (background). 
The time-dependent decay rate asymmetry $[N (\deltat) - \overline{N}(\deltat) ] / [  N (\deltat) + \overline{N}(\deltat) ]$
is also shown, where $N$ $(\overline{N})$ is the decay rate for \Bz(\Bzb) tagged events.

\begin{table}[!ht]
\caption{Contributions to the systematic errors on the signal yield, \S, and \C, where the
signal yield errors are given as number of events. The total systematic
uncertainty is the quadratic sum of the individual contributions listed. Additional systematic 
uncertainties that are applied only to the branching fraction are discussed in the text.~\label{table:systematicerrors}}
\begin{center}
\begin{tabular}{l|c|c|c}\hline\noalign{\vskip2pt}
Contribution                    & Yield       & \S\                  & \C\                 \\[2pt]\hline\hline\noalign{\vskip2pt}
PDF parameterization            & $^{+0.5}_{-1.6}$   & $^{+0.010}_{-0.012}$ & $^{+0.002}_{-0.011}$\\
Boost and $z$-scale               & $\pm 1.1$          & $\pm${0.001}        & $\pm${0.002}        \\
Beam spot position              & ---                & $\pm${0.004}        & $\pm${0.002}       \\
Fit bias                        & $\pm$ 1.5          & $\pm${0.021}        & $\pm${0.014}       \\
\B\ background yields           & $\pm 1.2$          & $\pm 0.029$         & $\pm 0.013$ \\
\CP\ content of \B\ background  & $\pm 0.4$          & $\pm 0.002$         & $\pm$0.002         \\
Tag side interference           & ---                & $\pm${0.004}        & $\pm${0.014}       \\\hline
Total                           & $^{+2.3}_{-2.7}$   & $\pm 0.04$& $\pm 0.03$\\[3pt]\hline\hline
\end{tabular}
\end{center}
\end{table}

Table~\ref{table:systematicerrors} summarizes the systematic uncertainties on 
the signal yield, \S, and \C. 
These include the uncertainty due to the PDF parameterization (including the resolution function), 
evaluated by varying the signal and background PDF parameters within the uncertainties of 
their nominal values. The PDF parameter uncertainties are determined from MC samples of signal
and background events. The uncertainties associated with the 
Lorentz boost, the $z$-scale of the tracking system, and the event-by-event beam spot position
are found to be small. 
We determine the fit bias on signal parameters from ensembles of generated
experiments using signal MC simulated data, which is generated using
the {\tt GEANT4}-based~\cite{ref:geant} \babar\ MC simulation, embedded 
into MC samples of background generated from the likelihood. 
We apply corrections to account for the observed fit bias on the signal 
yield, \S, and \C\ of $-2.7$ events, $-0.034$, and $-0.022$, respectively.
The uncertainty coming from this correction is taken as half of the correction
added in quadrature with the error on the correction.
Most, but not all, of the inclusive 
charmonium final states that dominate the inclusive \B\ background are precisely known 
from previous measurements. Their yields are fixed in the fit. 
As a cross check, yields for the \B\ backgrounds are allowed 
to vary one at a time. The sum in quadrature of deviations from the nominal result is 
taken as a systematic uncertainty. 
In order to evaluate the uncertainty coming from \CP\ violation in the \B\ background,
where appropriate, we introduce non-zero \S\ and \C\ for each background 
in turn. The uncertainty due to \CP\ violation in \btojpsiks\ is determined by 
varying \S\ and
\C\ within current experimental limits~\cite{ref:babar2007,ref:belle2007}.  For \B\ background events 
decaying into final states with charm, we allow for a 20\% asymmetry, and we allow for
100\% asymmetries in all other \B\ backgrounds.
We study the possible interference between the 
suppressed $\bar b\to \bar u c \bar d$ amplitude with the favored $b\to c \bar u d$ amplitude 
for some tag-side $B$ decays~\cite{ref:dcsd}. 
Systematic uncertainties from the effect of mis-alignment of the vertex detector and the use of an
effective lifetime for inclusive \B\ and $\jpsi K^{*0}$ backgrounds are found to be negligible.
There are additional systematic uncertainties that contribute only to the branching fraction.  These come from
uncertainties for \piz\ meson reconstruction efficiency (3\%), the $\jpsitoll$ branching fractions (1.4\%), 
the number of \B\ meson pairs (1.1\%), and tracking efficiency (1.0\%).  We apply a correction
for charged particle identification efficiency ($-1.3 \pm 0.7\%$ for $\jpsi\to e^+e^-$, 
and $-3.3 \pm 1.0\%$ for $\jpsi\to \mu^+\mu^-$ decays) based on the results of control sample studies using 
$\B$ decays with $\jpsi$ mesons in the final state. 
The systematic error contribution from MC statistics is negligible.  

We measure
\begin{eqnarray}
{\cal{B}} &=& \measuredbf,\nonumber\\
\S\ &=& \measuredS, \nonumber\\
\C\ &=& \measuredC, \nonumber 
\end{eqnarray}
where the correlation between \S\ and \C\ is \rhoSC. 
We determine the significance, including systematic uncertainties, of non-zero values of \S\ and \C\ using 
ensembles of MC simulated experiments as outlined in Ref.~\cite{ref:bellepipi}.
The significance of \S\ and \C\ being non-zero is $4.0\sigma$, which constitutes 
evidence for \CP\ violation in \btojpsipi\ decays.
The numerical values of \S\ and \C\ are consistent with the SM expectations for a 
tree-dominated ${b \rightarrow c\mskip 2mu \overline c \mskip 2mu d}$ transition. 
All results presented here are consistent
with previous measurements~\cite{ref:babarjpsipiz,ref:bellebf,ref:bellecp}.

We are grateful for the excellent luminosity and machine conditions
provided by our \pep2\ colleagues, 
and for the substantial dedicated effort from
the computing organizations that support \babar.
The collaborating institutions wish to thank 
SLAC for its support and kind hospitality. 
This work is supported by
DOE
and NSF (USA),
NSERC (Canada),
CEA and
CNRS-IN2P3
(France),
BMBF and DFG
(Germany),
INFN (Italy),
FOM (The Netherlands),
NFR (Norway),
MES (Russia),
MEC (Spain), and
STFC (United Kingdom). 
Individuals have received support from the
Marie Curie EIF (European Union) and
the A.~P.~Sloan Foundation.




\begin{thebibliography}{99}

\bibitem{babar-stwob-prl}
\babar\ Collaboration, B.\ Aubert {\em et al.},
\jprl{89}, 201802 (2002).

\bibitem{belle-stwob-prl}
Belle Collaboration, K.\ Abe {\em et al.},
\jprd{66}, 071102 (2002).

\bibitem{ref:CKM}
N.~Cabibbo, Phys.~Rev.~Lett.~{\bf 10}, 531 (1963);
M.~Kobayashi and T.~Maskawa, Prog.\ Th.\ Phys.\ {\bf 49}, 652 (1973).

\bibitem{BCP}
A.B.~Carter and A.I.~Sanda, \pr {\bf D23}, 1567 (1981);
I.I.~Bigi and A.I.~Sanda, \np {\bf B193}, 85 (1981).

\bibitem{grossman}
Y.~Grossman and M.~Worah,
\plb{395}, 241 (1997).

\bibitem{ciuchini}
M.~Ciuchini, M.~Pierini and L.~Silvestrini,
\jprl{95}, 221804 (2005).

\bibitem{ref:babar}
\babar\ Collaboration, B.\ Aubert {\em et al.},
\nima{479}, 1 (2002).

\bibitem{ref:pepii}
PEP-II: An Asymmetric B Factory. Conceptual Design Report, SLAC-R-418 (1993).

\bibitem{ref:babarjpsipiz}
\babar\ Collaboration, B.\ Aubert {\em et al.},
\jprd{74}, 011101 (2006).

\bibitem{ref:bellebf}
Belle Collaboration, K.\ Abe {\em et al.},
\jprd{67}, 032003 (2002).

\bibitem{ref:bellecp}
Belle Collaboration, K.\ Abe {\em et al.},
submitted to Phys. Rev. D. (Rapid Comm.), arXiv:0708.0304.

\bibitem{ref:bigprd}
\babar\ Collaboration, B.\ Aubert {\em et al.},
\jprd{66}, 032003 (2002).

\bibitem{ref:pdg2006}
Particle Data Group, 
W.~M.~Yao {\em et al.},\jpg{33}, 1 (2006), with partial update online.

\bibitem{ref:babar2007}
\babar\ Collaboration, B.\ Aubert {\em et al.},
\jprl{99}, 171803 (2007).

\bibitem{ref:hfag}
Heavy Flavour Averaging Group, E.~Barberio {\em et al.}, arXiv:0704.3575, with partial update online.

\bibitem{ref:argus}
The ARGUS Collaboration, H.\ Albrecht {\em et al.},
\plb{241}, 278 (1990).

\bibitem{babar:rhorhoprd}
\babar\ Collaboration, B.\ Aubert {\em et al.},
\jprd{76}, 052007 (2007).

\bibitem{ref:geant}
{\tt GEANT4} Collaboration, S.~Agostinelli {\em et al.},
\nima{506}, 250 (2003).

\bibitem{ref:belle2007}
Belle Collaboration, K.\ Abe {\em et al.}, 
\jprl{98}, 031802 (2007).

\bibitem{ref:dcsd}
O.~Long, M.~Baak, R.~N.~Cahn, and D.~Kirkby, 
\jprd{68}, 034010 (2003).

\bibitem{ref:bellepipi}
Belle Collaboration, K.\ Abe {\em et al.},
\jprd{68}, 012001 (2003).

\end{thebibliography}
\end{document}